\documentclass[conference]{IEEEtran}

\usepackage{cite}
\usepackage{amsmath,amssymb,amsfonts,dsfont}
\usepackage{algorithmic}
\usepackage{graphicx}
\usepackage{textcomp}
\usepackage{xcolor}
\usepackage[bookmarks=false]{hyperref}
\hypersetup{
     colorlinks=true,
     linkcolor=blue,
     filecolor=blue,
     citecolor=black,      
     urlcolor=cyan,
     }
\usepackage{url}
\def\BibTeX{{\rm B\kern-.05em{\sc i\kern-.025em b}\kern-.08em
    T\kern-.1667em\lower.7ex\hbox{E}\kern-.125emX}}
\begin{document}

\title{
Combining Weather Forecast Aggregation and State-Space Models for Adaptive Probabilistic Electricity Load Forecasting
}

\author{\IEEEauthorblockN{Joseph de Vilmarest}
\IEEEauthorblockA{
\textit{Viking Conseil}\\
Chartres, France}
\and
\IEEEauthorblockN{Jonathan Dumas}
\IEEEauthorblockA{
\textit{Réseau de Transport d'Électricité}\\
Paris, France}
\and
\IEEEauthorblockN{Jean Thorey}
\IEEEauthorblockA{
\textit{Enedis}\\
Paris, France}
}

\maketitle

\begin{abstract}

Accurate electricity load forecasting is essential to ensure the real-time balance between supply and demand, especially in systems increasingly influenced by weather conditions and renewable energy integration. In this paper, we propose an adaptive probabilistic forecasting framework that leverages multiple meteorological forecast providers to mitigate the impact of weather prediction errors on load forecasts. Our approach combines an adaptive generalized additive model (GAM) with a state-space formulation estimated via the Viking algorithm, enabling dynamic parameter updates over time. To improve uncertainty quantification, we extend the framework with a probabilistic layer based on GAM for location, scale and shape (GAMLSS), allowing the residual variance to depend on explanatory variables. We further introduce a short-term correction mechanism to enhance performance across different forecast horizons. A key contribution of this work is the aggregation of multiple meteorological inputs, whose diversity is shown to improve both point and probabilistic forecasts, particularly in capturing extreme quantiles by incorporating forecast dispersion. The methodology is evaluated on French national electricity consumption data and demonstrates consistent gains over single-provider approaches, with significant improvements in probabilistic forecasting accuracy.
\end{abstract}

\begin{IEEEkeywords}
Load forecasting, meteorological forecasts, time series.
\end{IEEEkeywords}

\section{Introduction}
Accurate electricity load forecasting is a cornerstone of power system operations, ensuring a continuous balance between supply and demand. This task is particularly critical for transmission system operators (TSOs), who rely on reliable forecasts to schedule generation, manage reserves, and guarantee system stability. In recent years, the increasing penetration of renewable energy sources, the electrification of end uses, and structural disruptions have made electricity demand more volatile and less stationary, thereby increasing the complexity of the forecasting problem.

Among the many drivers of electricity consumption, meteorological conditions play a central role, especially in countries such as France, where heating and cooling contribute significantly to demand variability. As a result, load forecasting models typically incorporate weather forecasts as key explanatory variables. 
However, inaccuracies in meteorological predictions directly propagate to load forecasts, making their reliability strongly dependent on the quality of weather inputs. In parallel, it is well established that combining multiple forecasts can improve predictive performance by exploiting the diversity of model errors \cite{gaillard2015contributions}. This idea has been extensively studied in the context of model aggregation. Yet, it has been less explored from the perspective of aggregating explanatory variables, such as meteorological forecasts, within a unified forecasting framework. Leveraging multiple weather forecast providers enables reducing the impact of individual forecast errors and obtaining more robust load predictions.

In this paper, we propose an adaptive probabilistic framework for electricity load forecasting that explicitly exploits the diversity of multiple meteorological forecast providers. Our approach is based on generalized additive model (GAM) \cite{wood2006generalized}, modelling the load as a sum of nonlinear effects of calendar and meteorological variables. 
The adaptive variant is obtained by a state-space formulation \cite{obst2021adaptive,de2022state}, enabling the model to dynamically adapt to evolving conditions. The intuition of adaptive algorithms stems from the rapid evolution of the electricity network ({\it e.g.} crises in recent years, electrification of uses, increase of renewable energies).
We extend this point-forecasting framework to a probabilistic setting by applying a GAMLSS-based correction \cite{stasinopoulos2017flexible}, thereby enabling a more accurate representation of forecast uncertainty. In addition, we introduce a short-term correction mechanism to improve forecasts across different horizons.

\begin{figure}[]
\centerline{\includegraphics[width=9cm]{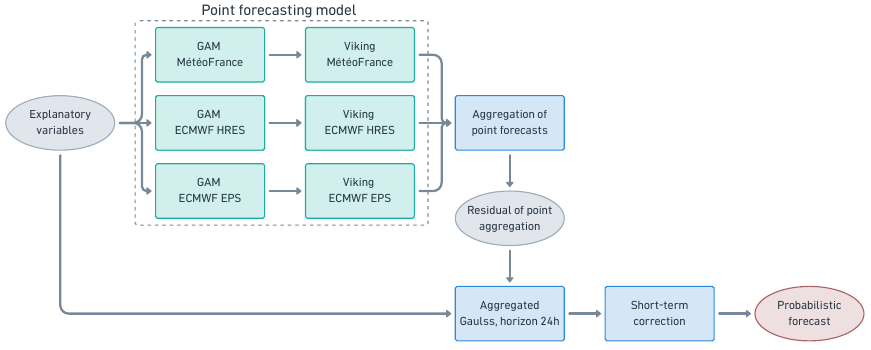}}
\caption{Probabilistic load forecasting approach using several weather forecast providers. %
}
\label{fig:graphical-abstract}
\end{figure}
The main contribution of this work, depicted by Figure \ref{fig:graphical-abstract}, lies in aggregating multiple meteorological inputs and integrating their dispersion into the probabilistic modeling stage. This not only improves point forecast accuracy but also better captures uncertainty, particularly at extreme quantiles. We evaluate the proposed methodology on the French national electricity load and demonstrate consistent improvements over single-provider approaches in both point and probabilistic forecasting tasks.

\section{Industrial Setting}\label{sec:data}
\begin{figure}[t]
\centering
\includegraphics[width=6.5cm]{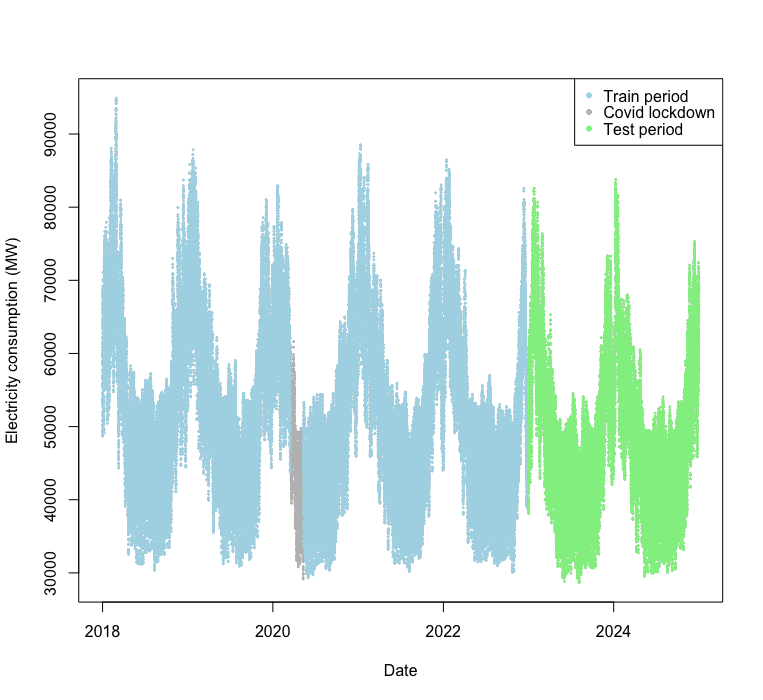}
\caption{Evolution of the French electricity consumption over time.}
\label{fig:load}
\end{figure}

%
%
The goal of this work is to predict the national electricity load in France. Figure \ref{fig:load} illustrates this target over time, as published by the French TSO: {\it Réseau de Transport d'Électricité}.
The forecast horizon is crucial in load forecasting because recent data can enhance initial predictions. Electricity suppliers are primarily interested in day-ahead forecasts, as these directly influence the spot market. Conversely, the TSO aims to maintain network balancing over time, which requires forecasts across various time horizons, particularly very short-term ones (ranging from a few minutes to a few hours in advance) to make optimal decisions in the balancing mechanism\footnote{\url{https://www.services-rte.com/en/learn-more-about-our-services/overview-of-market-mechanisms-managed-by-rte.html}}. In this work, we take into account the horizon of forecasts in two steps. First, we focus on forecasts made 24 hours in advance. Then, we apply a short-term correction, using the last available realized value to modify our forecasts over the different horizons.

\begin{figure}[t]
\centering
\includegraphics[width=4.35cm]{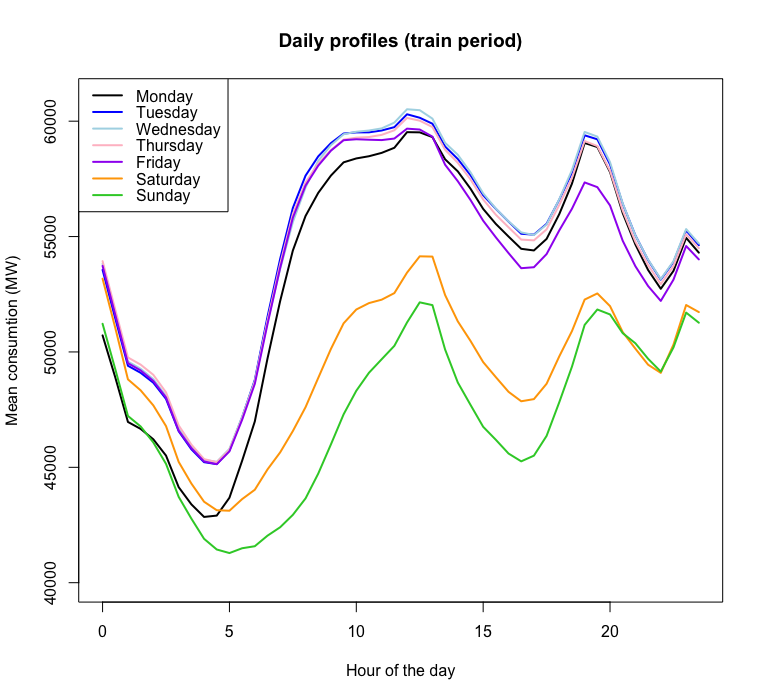}
\includegraphics[width=4.35cm]{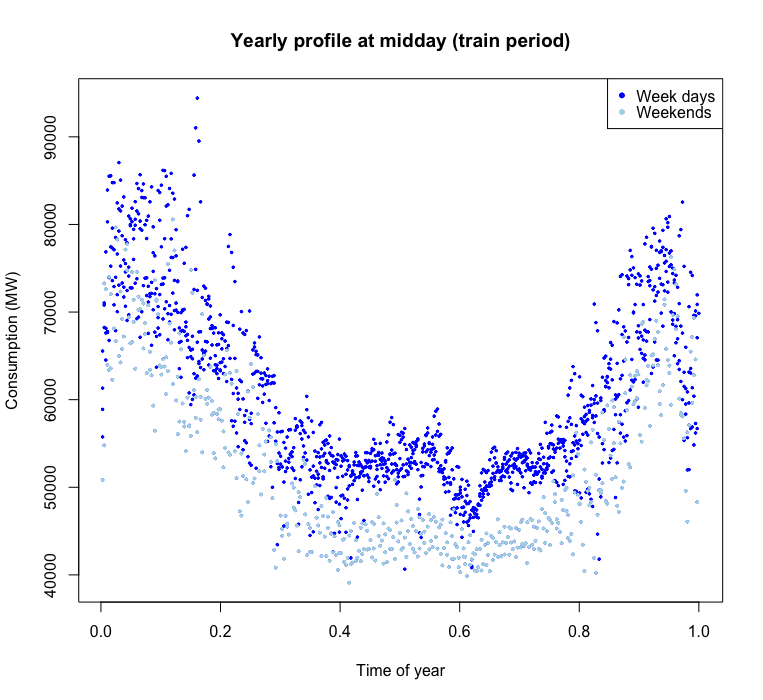}
\includegraphics[width=4.35cm]{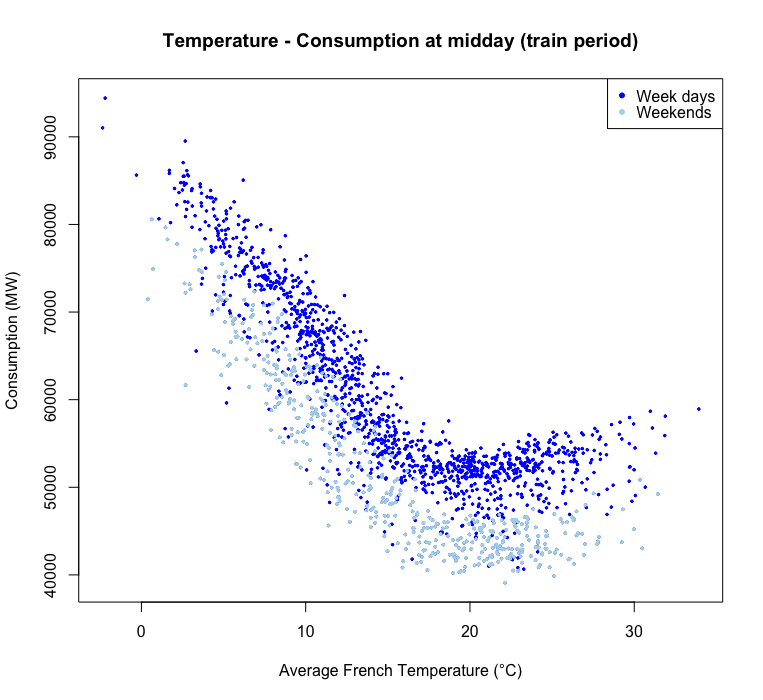}
\includegraphics[width=4.35cm]{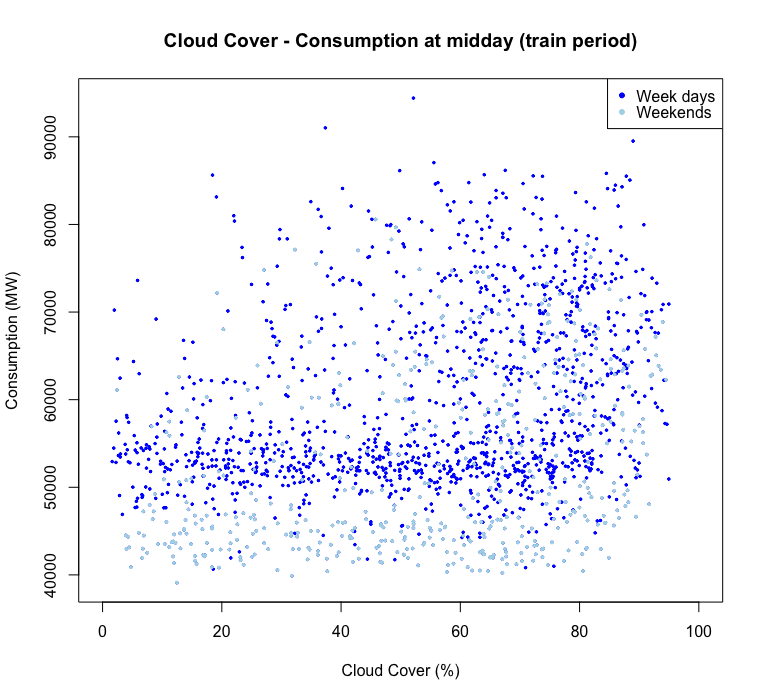}
\caption{Dependence of the consumption on calendar and meteorological variables.}
\label{fig:variable_load}
\end{figure}

We utilize calendar variables (such as time of day, day of the week, and holidays) along with meteorological variables (temperature and cloud cover forecasts) as explanatory variables. The relation between consumption and these covariates is illustrated in Figure~\ref{fig:variable_load}.
Consumption is lower on Saturdays compared to weekdays, and even lower on Sundays. Weekday consumption shows similar trends, with a noticeable decline on Monday mornings and Friday evenings. Throughout the year, consumption patterns are also distinct: it tends to be higher in winter than in summer, with a notable drop in activity during August. Temperature plays a significant role, as electric heating and cooling are commonly used in France. Additionally, the influence of cloud cover may vary, affecting both consumption and self-consumption.

\begin{figure}[t]
\centering
\includegraphics[width=4.35cm]{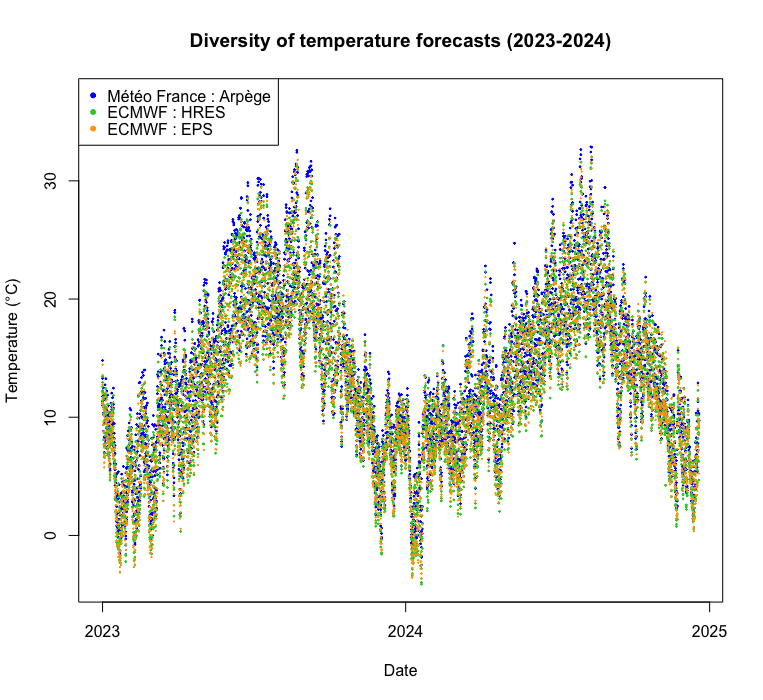}
\includegraphics[width=4.35cm]{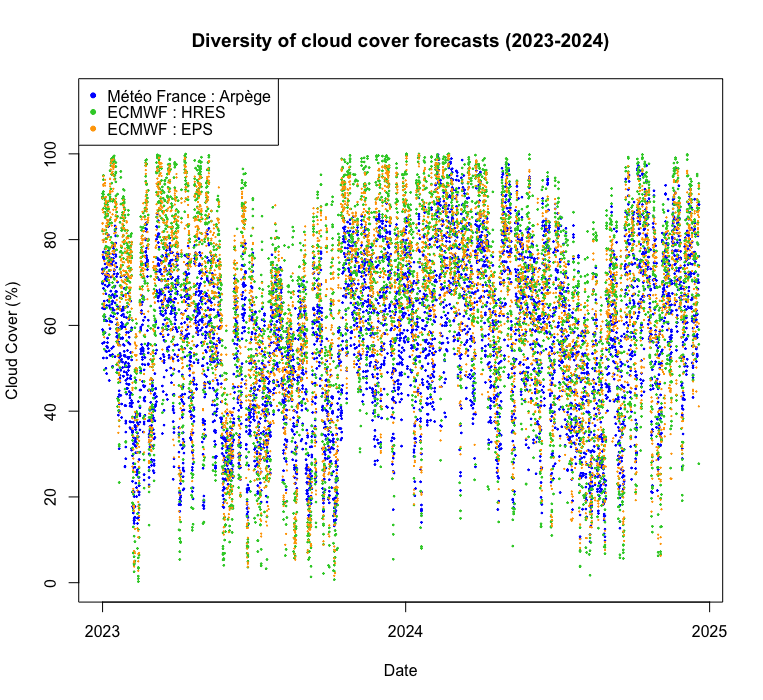}
\includegraphics[width=4.35cm]{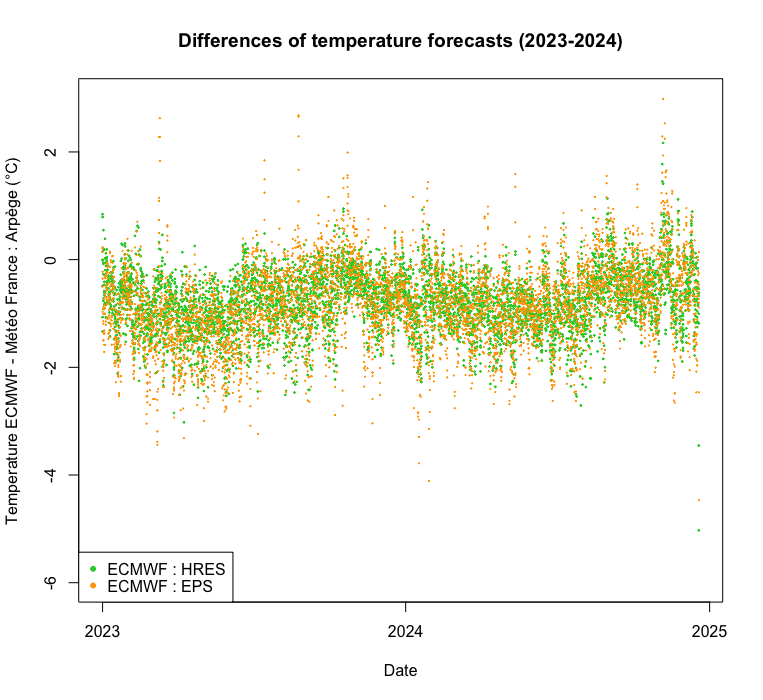}
\includegraphics[width=4.35cm]{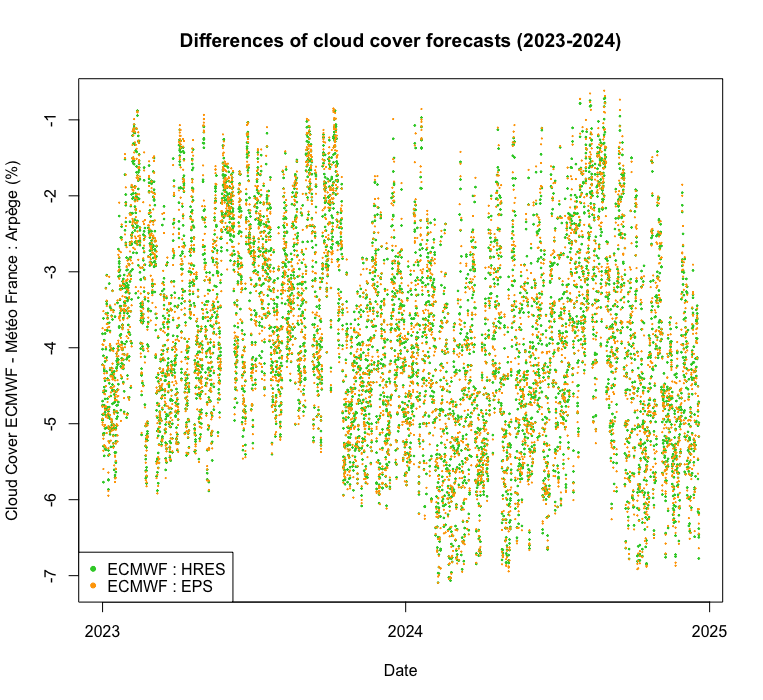}
\caption{Diversity of the meteorological forecasts.}
\label{fig:diversity_providers}
\end{figure}

%
The goal of our work is to utilize meteorological forecasts from multiple weather forecast providers. We have three distinct forecasts at our disposal, and their diversity is illustrated in Figure~\ref{fig:diversity_providers}.
%
%
We note that the three meteorological forecasts show significant differences, despite modeling a similar phenomenon. For instance, the standard deviations of their temperature differences are 0.57°C for Arpege-HRES, 0.74°C for Arpege-EPS, and 0.58°C for HRES-EPS.

\section{Adaptive Load Forecasting}\label{sec:methodology}
%
%
In this section, we outline our forecasting methodology, which depends solely on a single provider of meteorological forecasts. We denote the electricity load at time step $t$ as $y_t$, and the explanatory variables at this time step as $x_t$, where $x_t$ includes the meteorological forecasts from that one provider. In Section~\ref{sec:adaptive}, we summarize an existing adaptive algorithm. Then, in Section~\ref{sec:probabilistic}, we propose a probabilistic extension. For these first two steps we consider forecasts at 24-hour horizon, that is we forecast day $d$ hour $h$ having access to the realized target of day $d-1$ hour $h$. We evaluate in this setting in Section~\ref{sec:evaluation}. Finally, we apply a short-term correction in Section~\ref{sec:shortterm} in order to improve our forecasts for horizons different from 24 hours.

\subsection{Adaptive Generalized Additive Model}\label{sec:adaptive}
We utilize a forecasting approach that won a recent competition in post-COVID electricity load forecasting \cite{de2022state}. We employ a Gaussian GAM, based on the following equation:
\begin{align*}
	y_t = \sum\limits_{j=1}^d f_j(x_t) + \varepsilon_t\,,
\end{align*}
where each $f_j$ is a linear or non-linear function of one or two coordinates of $x_t$, and $\varepsilon_t$ is a Gaussian white noise.

In our application, the GAM effects are the following:
\begin{itemize}
\item
Smoothed effect of the {\it time of year} by a boolean {\it is week-end}, with cyclic cubic splines (continuity of the effect from December 31\textsuperscript{st} to January 1\textsuperscript{st});
\item
Categorical effect of the {\it day of week} by a boolean {\it is daylight saving time};
\item
Categorical effect of the {\it day of week} by a boolean {\it is bank holiday};
\item
Categorical effect of the {\it day of week} by a boolean {\it is bridge};
\item
Categorical effect of the {\it school holiday} (no holiday, Christmas holiday, summer holiday...);
\item
Categorical effect of a 3-category variable {\it tempo}, denoting a price signal for consumers;
\item
Smoothed effect of the {\it trend} (number of days since January 1\textsuperscript{st} 2018);
\item
Tensor effect of the {\it trend} and the {\it temperature};
\item
Smoothed effects of two {\it exponential moving averages of the temperature} of parameters $\sqrt{0.99}$ and $\sqrt{0.9}$;
\item
Linear effect of the {\it cloud cover};
\item
An intercept term.
\end{itemize}
We define $f(x_t)$ as the vector whose $j$\textsuperscript{th} coordinate is $f_j(x_t)$. 

We then obtain an adaptive variant by considering the following state-space model:
\begin{align*}
	& y_t - \theta_t^\top f(x_t) \sim \mathcal{N}(0,\sigma_t^2) \,,\\
	& \theta_{t+1} - \theta_t \sim \mathcal{N}(0, Q_t)\,.
\end{align*}
If the variances are predefined, the estimation of the state vector $\theta_t$ can be achieved using the Kalman filter \cite{kalman1960new}. In situations with unknown variances, various methods have been developed. Following the approach outlined by \cite{de2022state}, we utilize the Viking algorithm \cite{vilmarest2024viking}. This algorithm introduces additional latent variables within an augmented state-space model. In our case, we fix $\sigma_t^2 = \sigma^2$ and adjust only the covariance matrix $Q_t$, based on the following considerations:
\begin{subequations}
\begin{align}
	\label{eq:observation}& y_t - \theta_t^\top f(x_t) \sim \mathcal{N}(0,\sigma^2) \,,\\
	\label{eq:state}& \theta_{t+1} - \theta_t \sim \mathcal{N}(0, h(b_t))\,, \\
	\label{eq:b}& b_{t+1}-b_t \sim \mathcal{N}(0, \rho_b I)\,,
\end{align}
\end{subequations}
where $h$ is the default pre-defined function \cite{vilmarest2024viking} and the parameter $\rho_b$ is optimized on a validation period.

\subsection{Probabilistic}\label{sec:probabilistic}

Like the Kalman filter, Viking already produces probabilistic forecasts. Indeed, the state estimation obtained from the Viking algorithm is represented as a Gaussian distribution:
\begin{align*}
	\theta_t\mid (f(x_s),y_s)_{s<t} \sim \mathcal{N}\big(\hat\theta_t, P_t\big)\,.
\end{align*}
%
%
Combining with Equation \eqref{eq:observation}, we obtain
\begin{align*}
& y_t\mid (f(x_s),y_s)_{s<t}, f(x_t) \sim\mathcal{N}\big(\hat y_t,\hat\sigma_t^2\big)\,,\\
& \hat y_t=\hat\theta_t^\top f(x_t)\,, \hat\sigma_t^2=\sigma^2+f(x_t)^\top P_t f(x_t)\,.
\end{align*}

%
Therefore, the forecasting method discussed in Section~\ref{sec:adaptive} provides a Gaussian probabilistic forecast. However, the quality of this forecast largely depends on how well the state-space model is specified. In practice, the assumption of a Gaussian distribution with fixed variance is often too rigid.

A way to relax this hypothesis is by allowing the variance to depend on covariates. This is achieved using GAMLSS \cite{stasinopoulos2017flexible}.
%
%
%
More specifically, we apply a Gaulss: we assume a Gaussian distribution for the residuals of the point forecasts and model their variance using an additive approach. Formally, we state that:
\begin{align*}
	& y_t-\hat y_t \mid x_t \sim \mathcal{N}(0, s_t^2)\,, \\
    & g(s_t) = \sum\limits_{j=1}^d f_j(x_t)\,,
\end{align*}
where $g:s\in\mathbb{R}_+\rightarrow \log(s-b)$ is the link function for some $b>0$ and each $f_j$ is a linear or non-linear function of one or two coordinates of $x_t$. We rely on the following effects:
\begin{itemize}
\item
Linear effects of $\log\hat\sigma_t$ and $\log\hat y_t$;
\item
Categorical effects of the boolean {\it is bank holiday} and of {\it school holiday};
\item
Smoothed effect of the {\it time of year}, with cyclic cubic splines;
\item
Smoothed effects of the {\it temperature} and of the {\it cloud cover}.
\end{itemize}

\subsection{Evaluation}\label{sec:evaluation}
\begin{figure}[t]
\centering
\includegraphics[width=4.35cm]{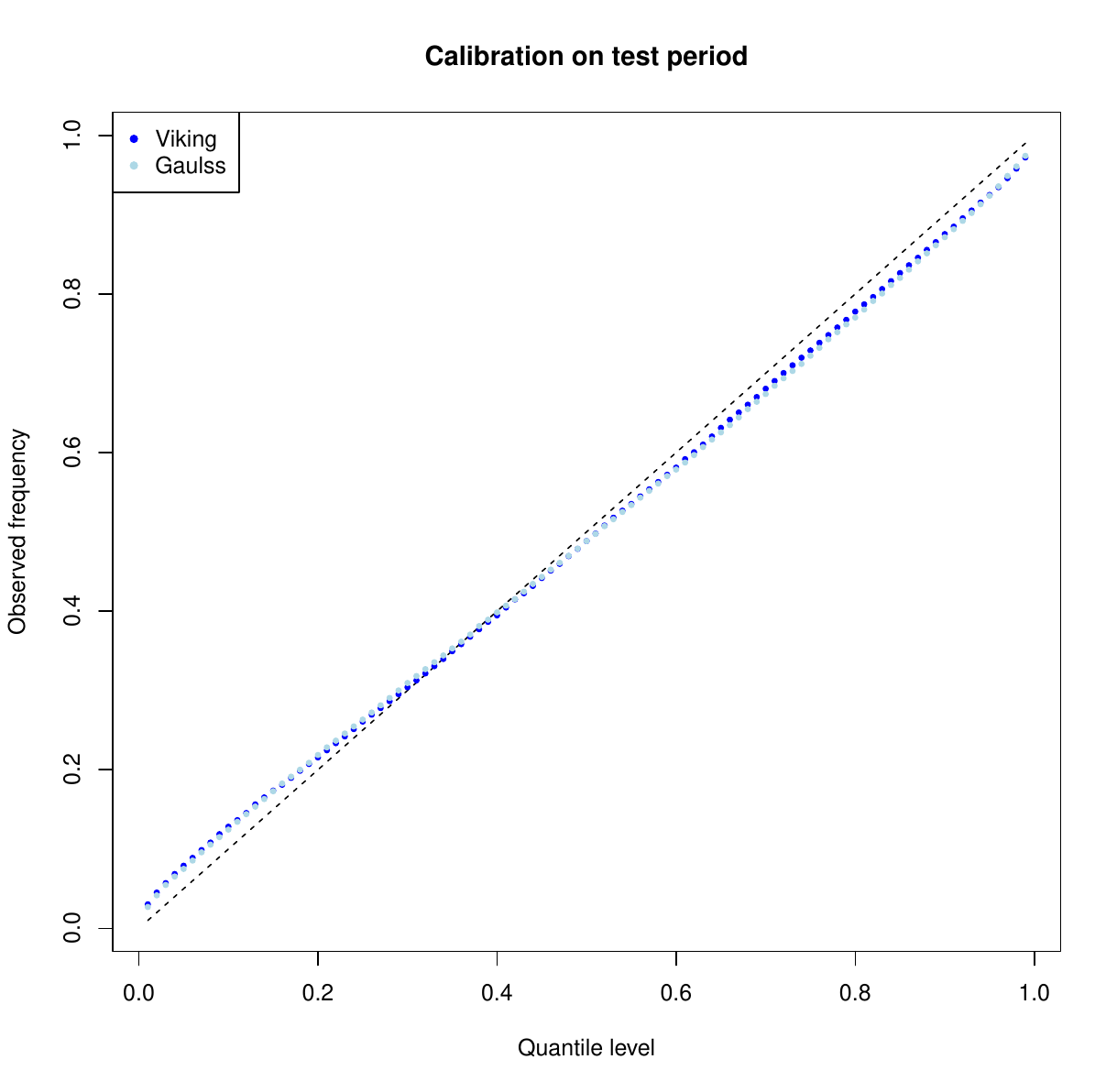}
\includegraphics[width=4.35cm]{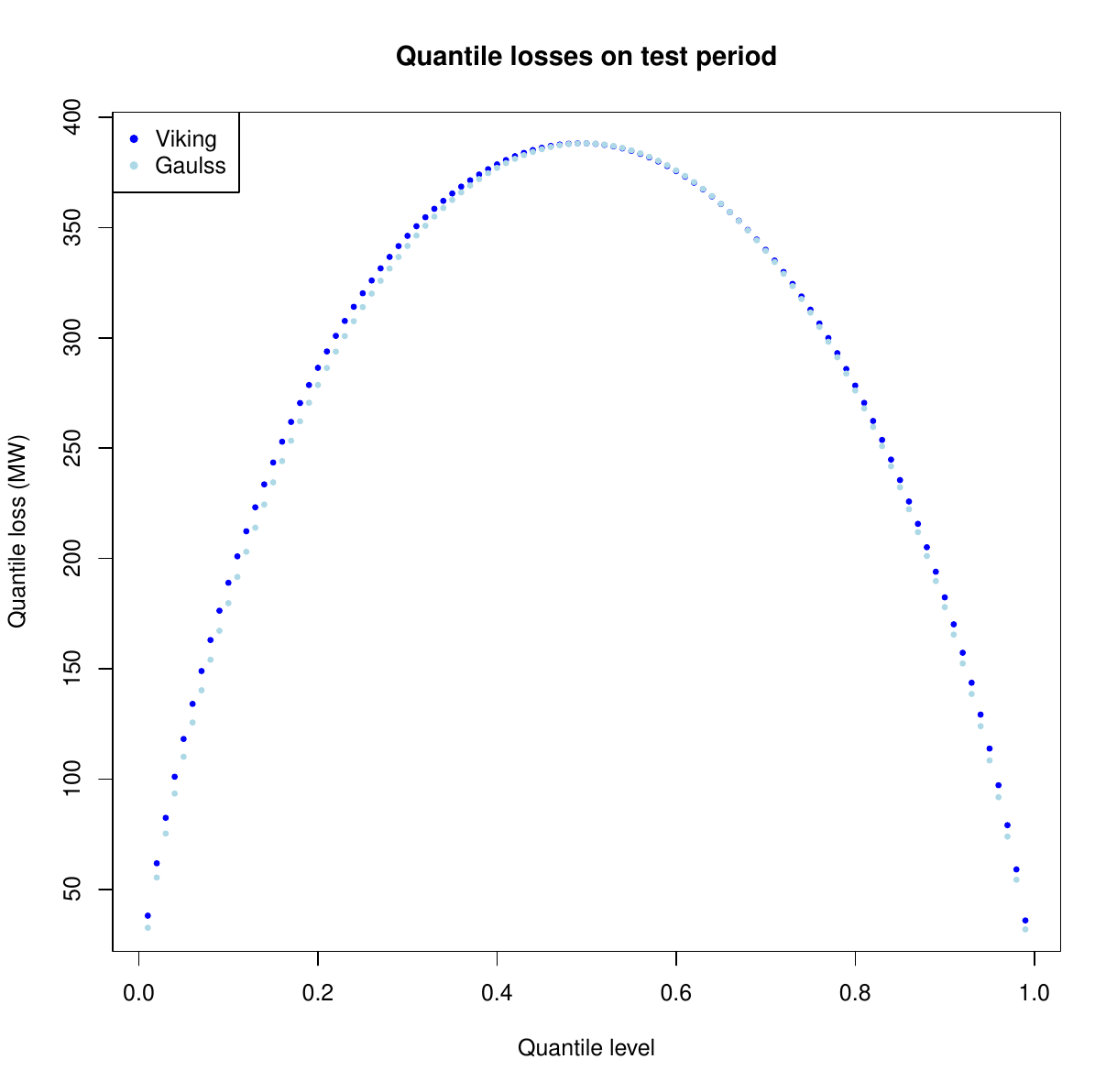}
\caption{Calibration (on the left) and quantile losses (on the right) of the two Gaussian probabilistic forecasts : GAM Viking, and the Gaulss correction.}
\label{fig:evaluation_probabilistic}
\end{figure}

To evaluate our probabilistic forecasts, we rely on the paradigm 
\begin{quote}
    {\it maximizing the sharpness of the predictive distributions subject to calibration} \cite{gneiting2007probabilistic}. 
\end{quote}
The sharpness is maximized by the most concentrated distribution, with the condition of calibration, or coherence between observed and forecasted distributions. We assess calibration by the coincidence between levels of quantiles and frequencies of exceedances.  
The calibration results are presented in Figure~\ref{fig:evaluation_probabilistic}, and our analysis indicates that the forecasts are well-calibrated.

As calibration is satisfied, the sharpness is evaluated by the {\it continuous ranked probability score} \cite{gneiting2007strictly}. We use the following discrete approximation of the CRPS, for a set of quantile forecasts $(\hat y_{q_1}\hdots \hat y_{q_l})$ of the observation $y$:
$$
CRPS((\hat y_{q_1}\hdots \hat y_{q_l}),y) = \sum\limits_{i=1}^l \rho_{q_i}(y, \hat y_{q_i}) (q_{i+1}-q_{i-1}) \,,
$$
where $q_0=0, q_{l+1}=1$, and $\rho_q$ is the quantile loss defined by
$$
\rho_q(y,\hat y_q) = (\mathds{1}_{y<\hat y_q}-q) (\hat y_q - y) \,.
$$
%
%
We use percentiles : $q_i=i/100$ for $1\le i\le 99$; then, the CRPS is averaged over the test period. The main drawback of this metric is that extreme quantiles are under-represented. Indeed, the obtained CRPS is 285 MW for GAM-Viking and 281 MW for GAM-Viking-Gaulss, that is a gain of $1.2\%$, whereas the gain is of $14\%$ (resp. $11\%$) on the quantile level $1\%$ (resp. $99\%$), {\it c.f.} Figure~\ref{fig:evaluation_probabilistic}.

\subsection{Short-term Correction}\label{sec:shortterm}
The methodology presented in Sections~\ref{sec:adaptive} and \ref{sec:probabilistic} is designed at 24-hour horizon. We introduce a {\it short-term correction} in order to take into account the horizon and improve our forecasts especially for shorter horizon. Indeed, when we forecast time step $t$ at time $t-h$ ($h$ is the forecasting horizon), we have access to the target $y_{t-h}$. We integrate this information through the error of our model.

\begin{figure}
\centering
\includegraphics[width=4.35cm]{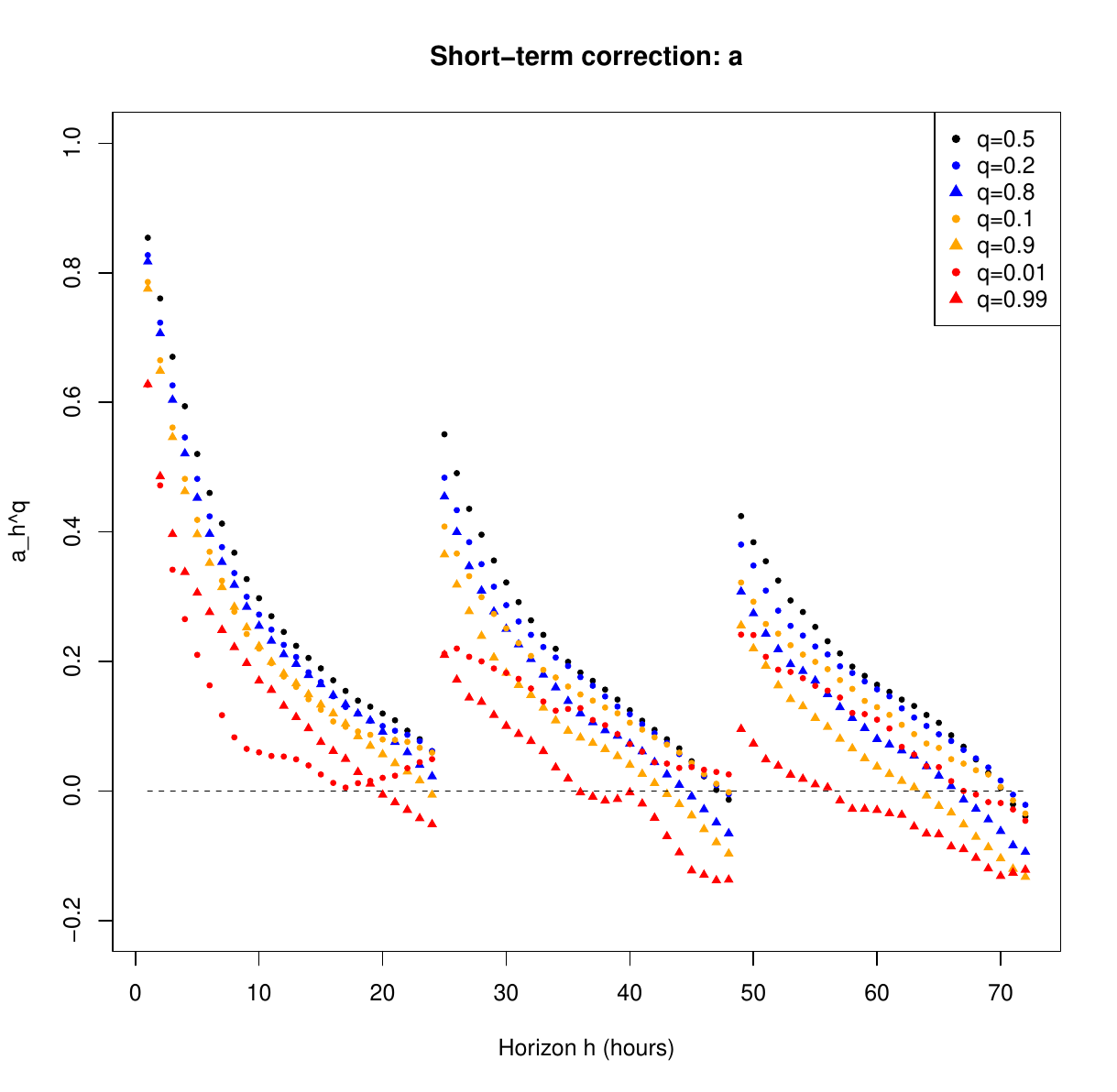}
\includegraphics[width=4.35cm]{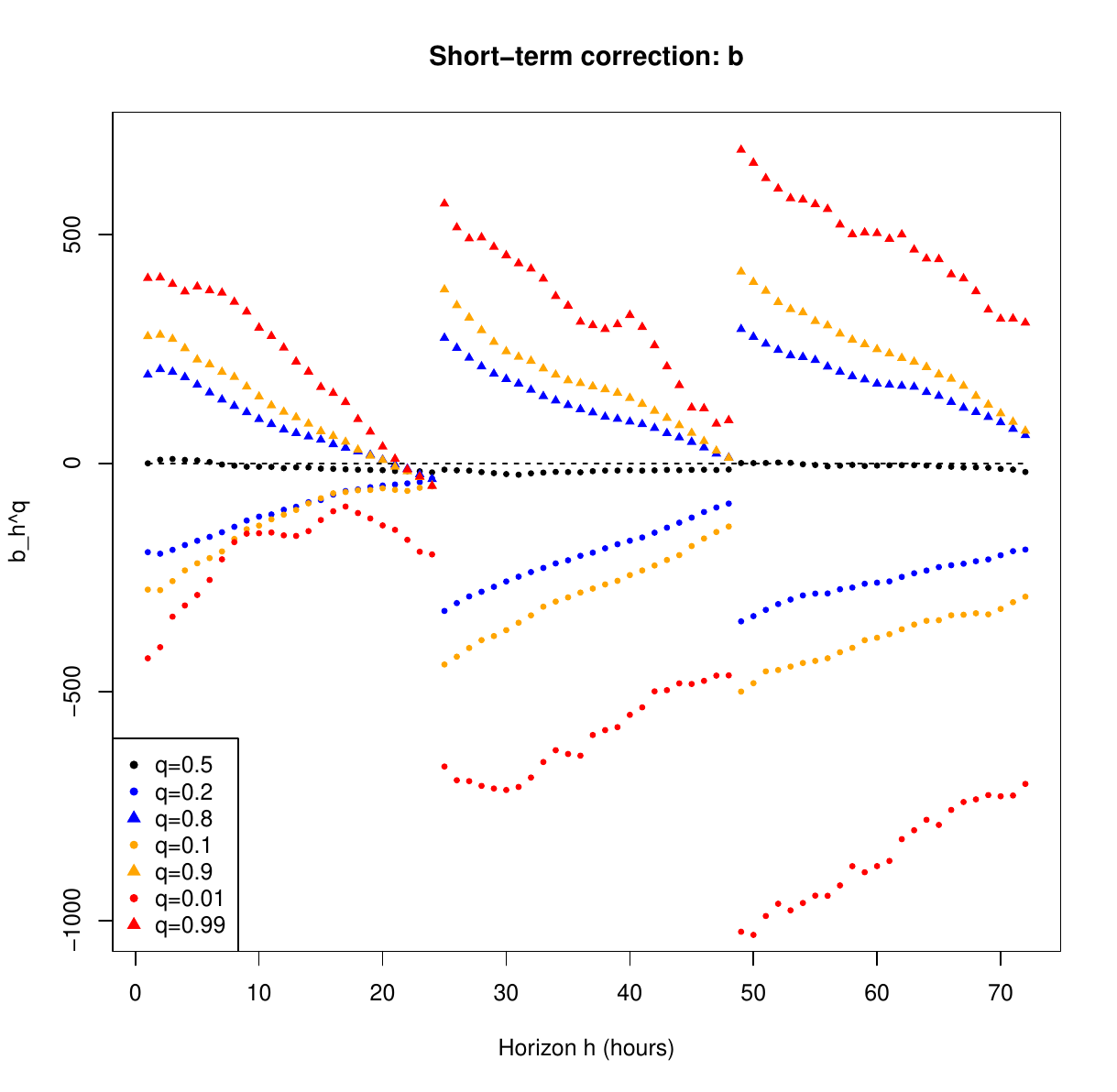}
\caption{Coefficients of the short-term correction as a function of horizon $h$, applied on the Gaulss forecasts for different quantile levels.}
\label{fig:coef_ct}
\end{figure}

Formally, for any quantile level $q$ and forecasting horizon $h$, we use the last quantile residual $y_{t-h}-\hat y_{t-h}^q$ to correct our initial forecast $\hat y_t^q$ defined in Section~\ref{sec:probabilistic}. We optimize $a_{h}^q,b_{h}^q$ in the following linear quantile regression:
$$
\hat y_t^{q,ST} = \hat y_t^q + a_{h}^q (y_{t-h}-\hat y_{t-h}^q)+b_{h}^q.
$$
The coefficients $a_{h}^q,b_{h}^q$ are obtained by quantile loss minimization on the train set. Obtained coefficients are displayed in Figure~\ref{fig:coef_ct}.
For very-short horizons (a few hours), the term $a_h^q (y_{t-h}-\hat y_{t-h}^q)$ is the important one because the autocorrelation of consecutive values of the target is high. For horizons longer than 24 hours, the second term $b_{h}^q$ is important to add a constant term increasing the dispersion of the forecast as the horizon (and thus uncertainty) increases.

\section{Aggregating Meteorological Forecasts}\label{sec:aggregation}
%
%
In this section, we combine multiple meteorological forecasts to leverage their complementary strengths, using the adaptive method outlined in Section~\ref{sec:adaptive} as the base model. We present  the improvements achieved both at the point level and in probabilistic forecasts.


\subsection{Impact on Point Forecasts}
We apply our adaptive algorithm to each meteorological provider in order to obtain the probabilistic prediction for each provider $m$, that we denote as $\mathcal{N}(\hat y_t^{(m)},\hat\sigma_t^{(m)})$.
To aggregate the multiple point forecasts, we rely on the weighted average:
\begin{align*}
	\hat y_t^{(agg)} = \sum\limits_m p_t^{(m)} \hat y_t^{(m)}\,,
\end{align*}
and we test various heuristics to define the weights $p_t^{(m)}$.

\subsubsection{Uniform weights}
this very simple method provides an interesting benchmark.
\subsubsection{Time-invariant weights}
\begin{figure}[t]
\centering
\includegraphics[width=4.35cm]{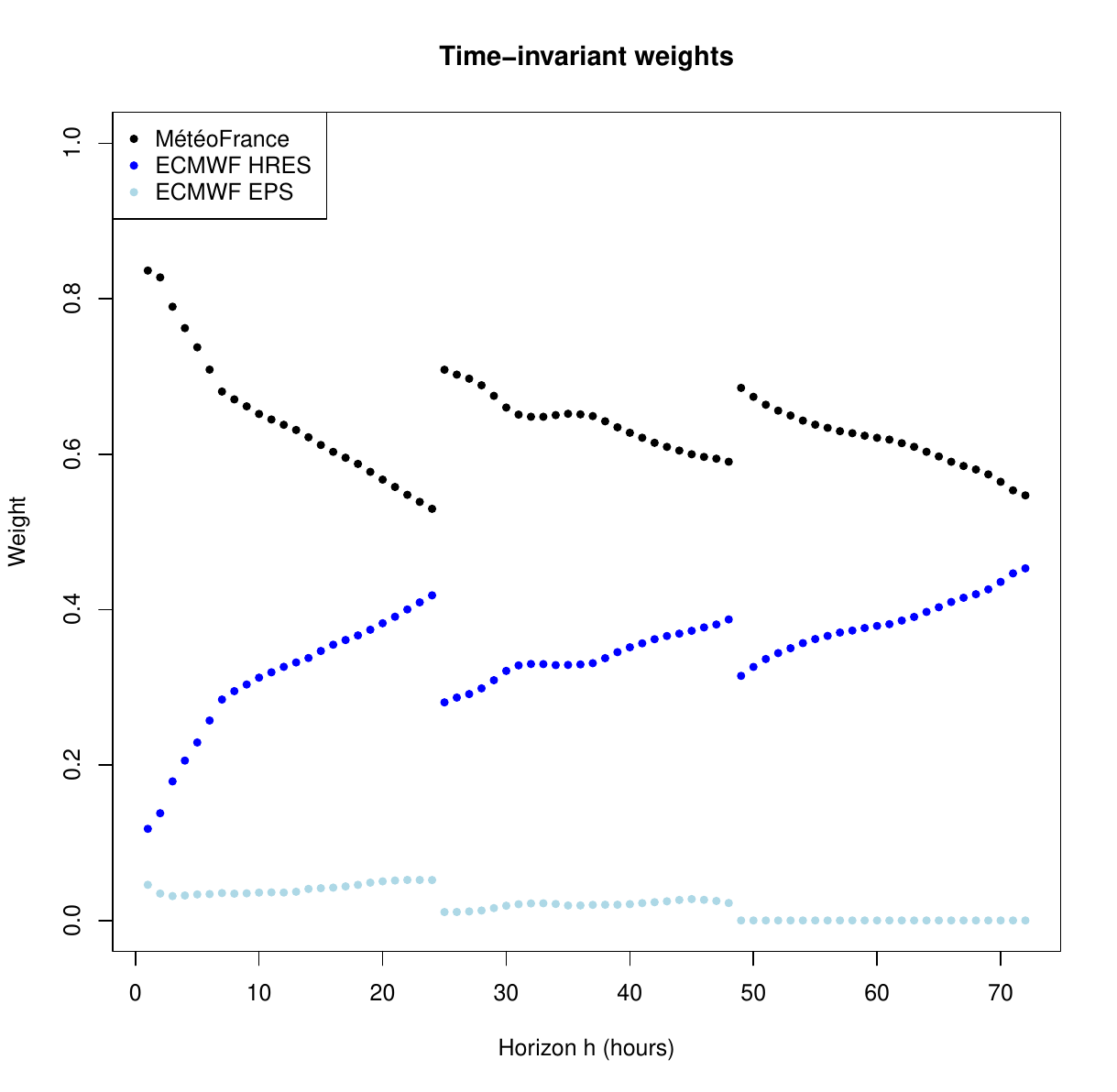}
\caption{Time-invariant weights attributed to each forecast as a function of the horizon.}
\label{fig:weights_agg_horizon}
\end{figure}

we optimize time-invariant weights $p^{(m)}$ on the validation period 2021-2022.
This optimization is realized by horizon, and we plot the obtained weights in Figure~\ref{fig:weights_agg_horizon}.
We choose to optimize over a shorter period than the training of our forecasting models because meteorological forecasts have changed over time, and the beginning of our data set is not representative of their current process.
\subsubsection{Sequential aggregation of experts}
we rely on the paradigm of expert's aggregation\cite{gaillard2015contributions} to estimate time-varying weights. We apply {\it Bernstein Online Aggregation} \cite{wintenberger2017optimal} with the squared loss.
\subsubsection{Additive stacking} 
\begin{figure}[t]
\centering
\includegraphics[width=4.35cm]{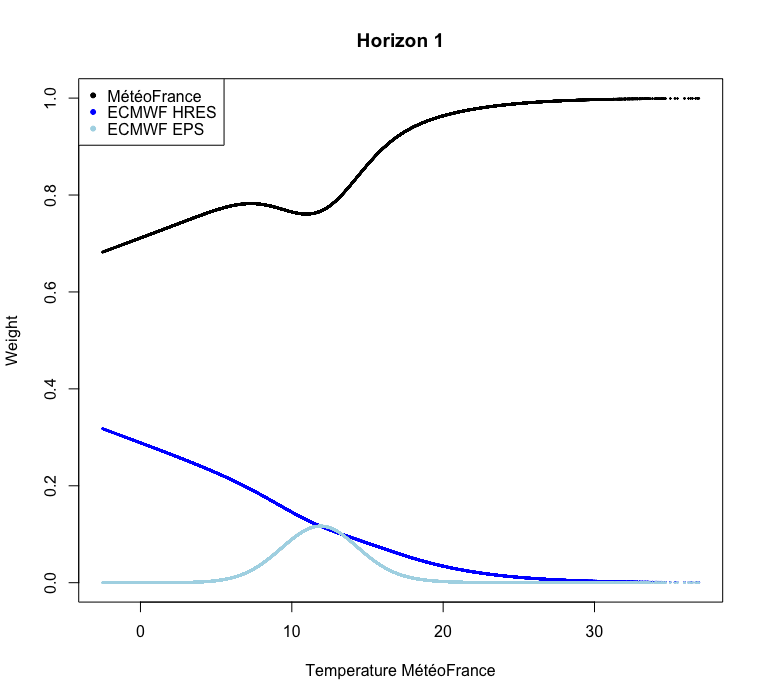}
\includegraphics[width=4.35cm]{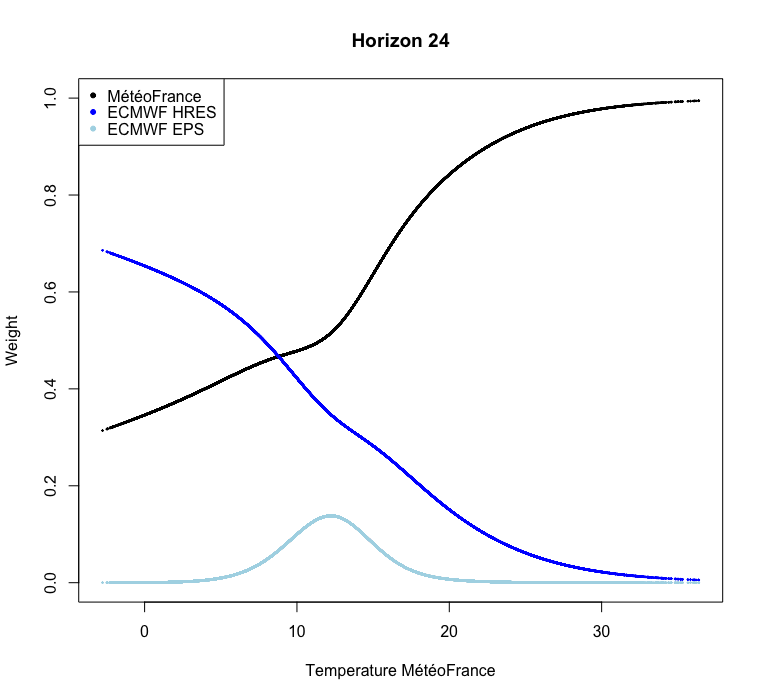}
\caption{Weights attributed to each forecast as a function of the MétéoFrance temperature forecast, for horizon 1 hour (left) and 24 hours (right). The forecasts of MétéoFrance are more reliable for hot temperatures, while the ones of ECMWF are better for cold temperatures.}
\label{fig:weights_gamfactory}
\end{figure}

we optimize weights that depend on the forecasted temperature (MétéoFrance); we rely on additive stacking implemented by the R package \texttt{gamFactory} \cite{capezza2021additive}. The obtained weights are represented in Figure \ref{fig:weights_gamfactory}.

We analyse the performance obtained by the point forecasts $(\hat y_t)_t$ in terms of the RMSE defined on period $\mathcal{T}$ by:
\begin{align*}
RMSE = \sqrt{\frac{1}{|\mathcal{T}|} \sum\limits_{t\in\mathcal{T}} (y_t-\hat y_t)^2}\,.
\end{align*}

\begin{figure}[t]
\centering
\includegraphics[width=4.35cm]{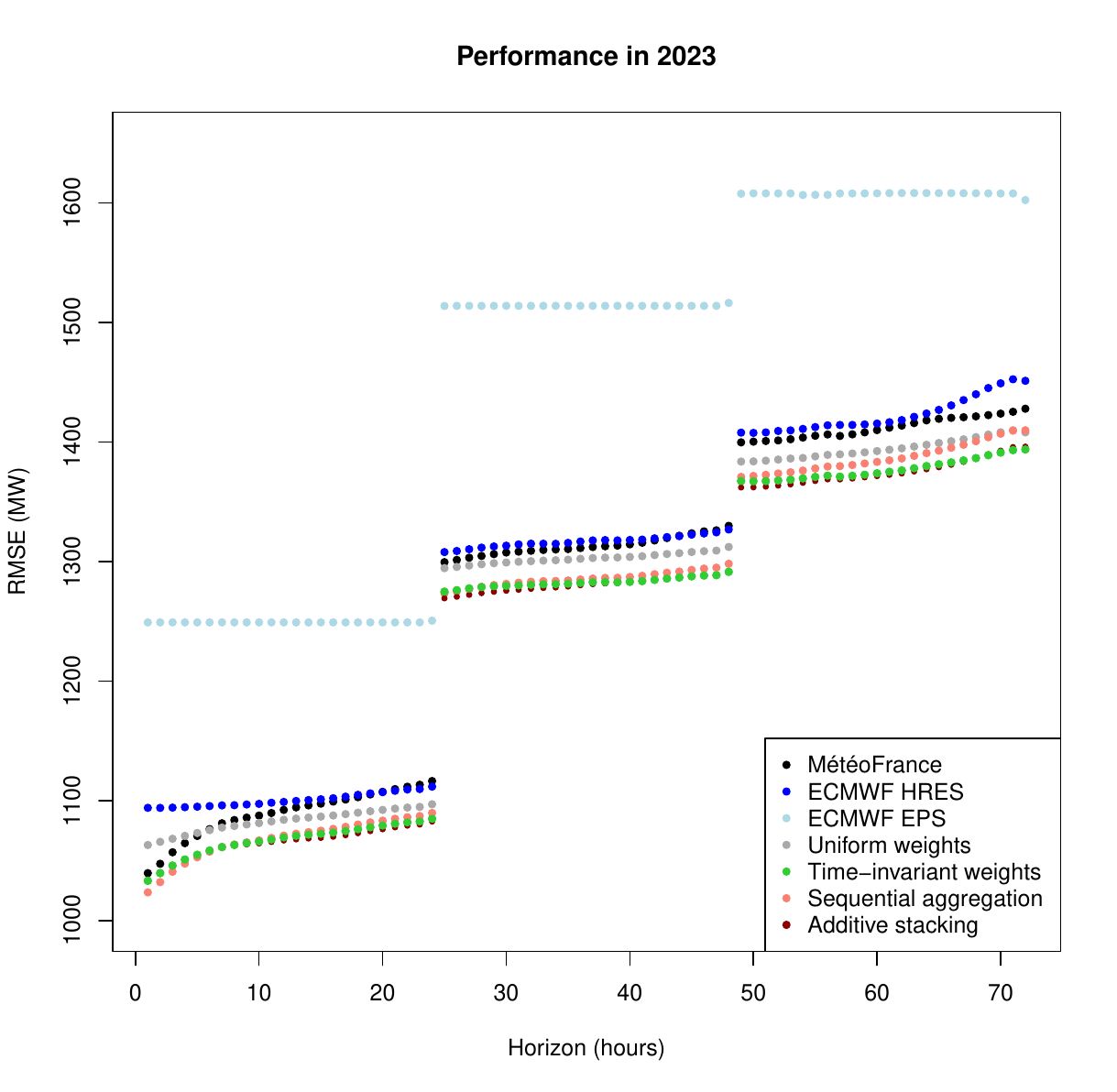}
\includegraphics[width=4.35cm]{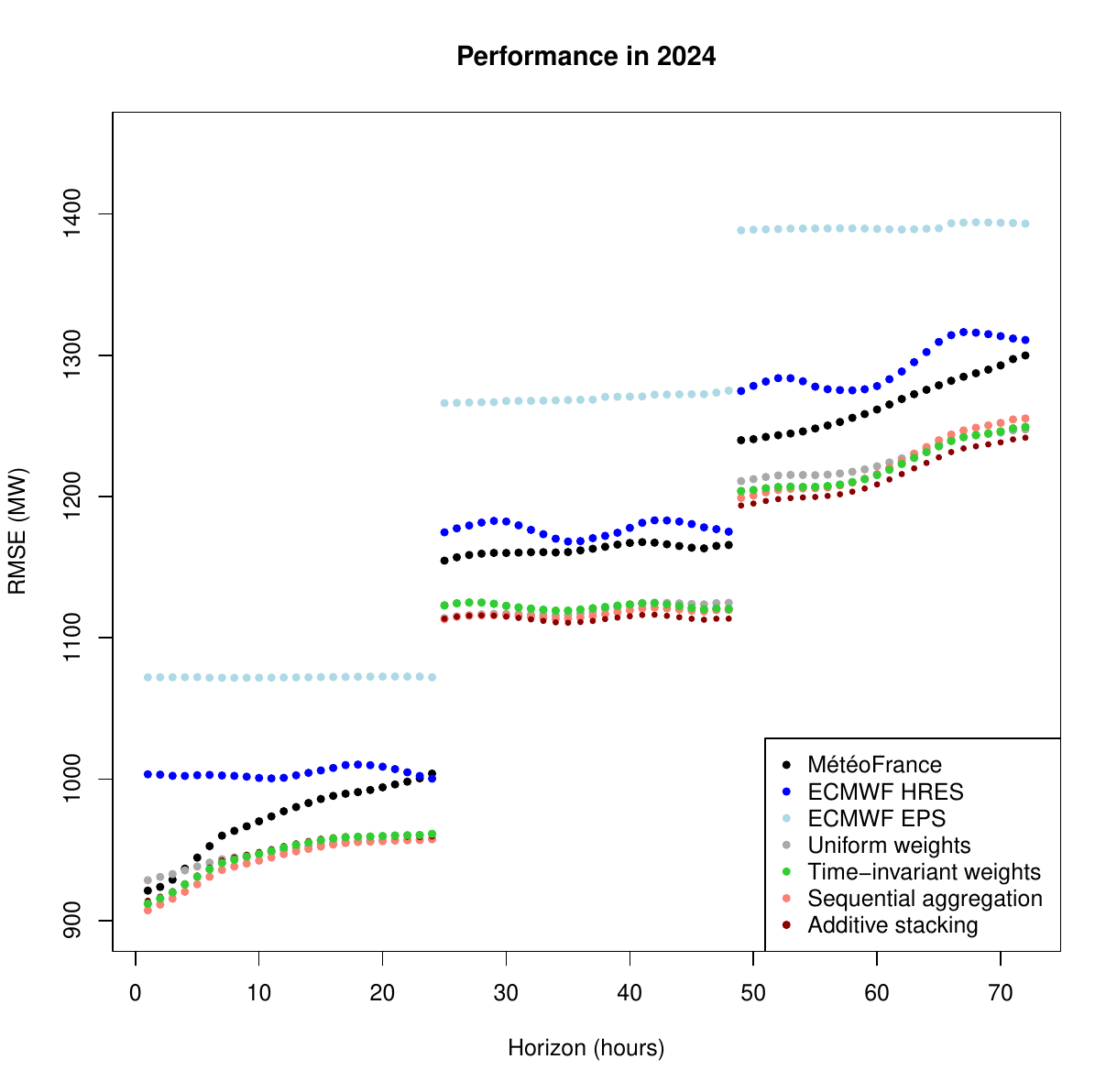}
\caption{RMSE obtained by each method as a function of the horizon.}
\label{fig:perf_agg_det}
\end{figure}

The performance of the different methods is displayed in Figure~\ref{fig:perf_agg_det} as a function of the horizon, and summarized in Table~\ref{tab:perf_point} at 24-hour horizon. For short horizons the MétéoFrance forecasts yield much better load forecasts, therefore the weight is higher for this one (Figures~\ref{fig:weights_agg_horizon} and \ref{fig:weights_gamfactory}). Notably, at 24-hour horizon, the various heuristics tested yield very similar results, and a uniform average already accounts for most of the gains. We decided to use time-invariant weights, which offer a slight improvement over uniform weights with no substantial addition of complexity.

\begin{table}[t]
\renewcommand{\arraystretch}{1.25}
\caption{Performances of the point forecasts obtained from each meteorological provider. RMSE in MW is provided for 2023 and 2024 at 24-hour horizon.}
\begin{center}
\begin{tabular}{|c|c|c|}
\hline
Provider & 2023 & 2024 \\
\hline
MétéoFrance : Arpège & 1117 MW & 1004 MW \\
ECMWF : HRES & 1112 MW & 1000 MW \\
ECMWF : EPS & 1251 MW & 1072 MW \\
\hline
Uniform & 1097 MW & 957 MW \\
\hline
Time-invariant & 1085 MW & 961 MW \\
\hline
Sequential aggregation & 1090 MW & 958 MW \\
\hline
Additive stacking & 1082 MW & 960 MW \\
\hline
\end{tabular}
\label{tab:perf_point}
\end{center}
\end{table}

\subsection{Impact on Probabilistic Forecasts}
%
%
Aggregating data from various weather providers can enhance the information available for probabilistic forecasts by providing a measure of forecast dispersion. 
As for the point forecasts, each meteorological provider yields a probabilistic forecast $\mathcal{N}(\hat y_t^{(m)}, \hat\sigma_t^{(m)})$ after the application of the Gaulss model on the residuals of the point forecast. We decide not to combine these distributions, but rather to aggregate the deterministic forecasts in a new Gaulss on an aggregated covariate vector. Formally, using $\hat y_t^{(agg)}$ and the corresponding standard deviation $\hat\sigma_t^{(agg)2} = \sum_m p_t^{(m)} \hat \sigma_t^{(m)2}$, we create a new Gaulss using as covariate vector the vector $z_t^{(agg)}$ where each variable is replaced by its aggregated version. We then apply the Gaulss defined in Section~\ref{sec:probabilistic}. We compare it to the aggregated Gaussian forecast $\mathcal{N}(\hat y_t^{(agg)}, \hat\sigma_t^{(agg)})$.

%
To that first aggregated version of Gaulss (Gaulss Agg 1), we compare several variants, with the intuition that multiple initial forecasts yield information about uncertainty through their dispersion. 
We add a new covariate in the Gaulss measuring the dispersion of the forecasts coming from the different meteorological providers, namely the standard deviation of $(\hat y_t^{(m)})_m$; that is Gaulss Agg 2. We then add also the standard deviation of $(\hat\sigma_t^{(m)})_m$; that is Gaulss Agg 3. Finally, instead of using the average standard deviation and its standard deviation, we use the three standard deviations as three covariates in the Gaulss; that is Gaulss Agg 4.

\begin{figure}[t]
\centering
\includegraphics[width=4.35cm]{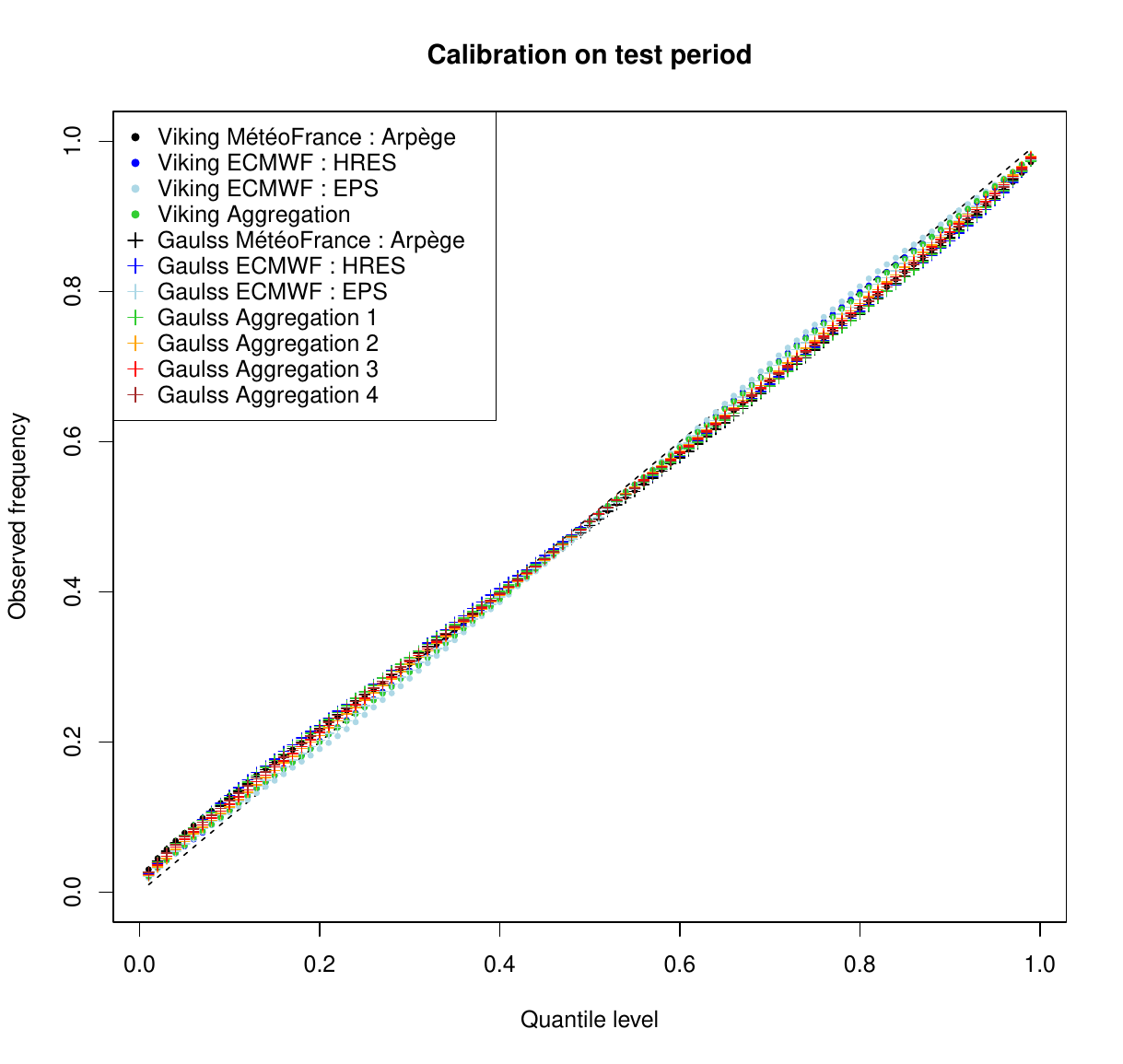}
\includegraphics[width=4.35cm]{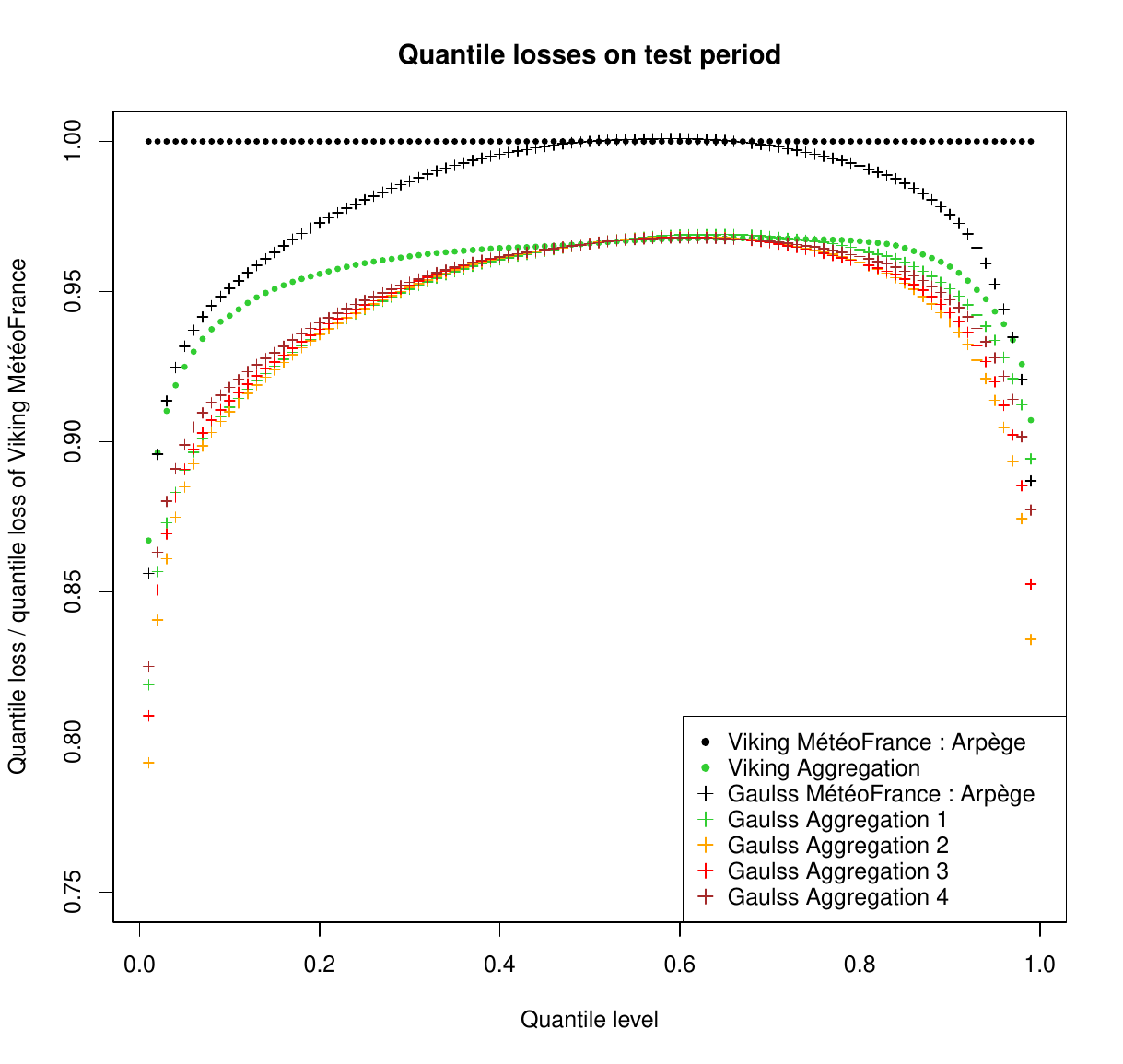}
\caption{Calibration (on the left) and quantile losses (on the right) of the different aggregated variants at 24-hour horizon.}
\label{fig:proba_agg}
\end{figure}

Figure~\ref{fig:proba_agg} presents the results at 24-hour horizon.
%
%
We observe that all methods are well-calibrated. The graph of the quantile losses reveals several insights. First, the aggregation method using GAM - Viking shows significant improvement at the median, which translates to the same benefit for the GAM - Viking - Gaulss variants, as their median forecasts are identical. In the tails, the latter method performs much better, as expected, and we note that the second variant produces the best results by incorporating information about the dispersion of the point forecasts.
Numerically, the gain of the Viking Aggregation over Viking MétéoFrance is of 13\% and 9\% for the quantiles 1\% and 99\%; for Gaulss Aggregation 1 (resp. Gaulss Aggregation 2) it is of 18\% and 11\% (resp. 21\% and 17\%) for the same quantiles. 

\begin{figure}[t]
\centering
\includegraphics[width=4.35cm]{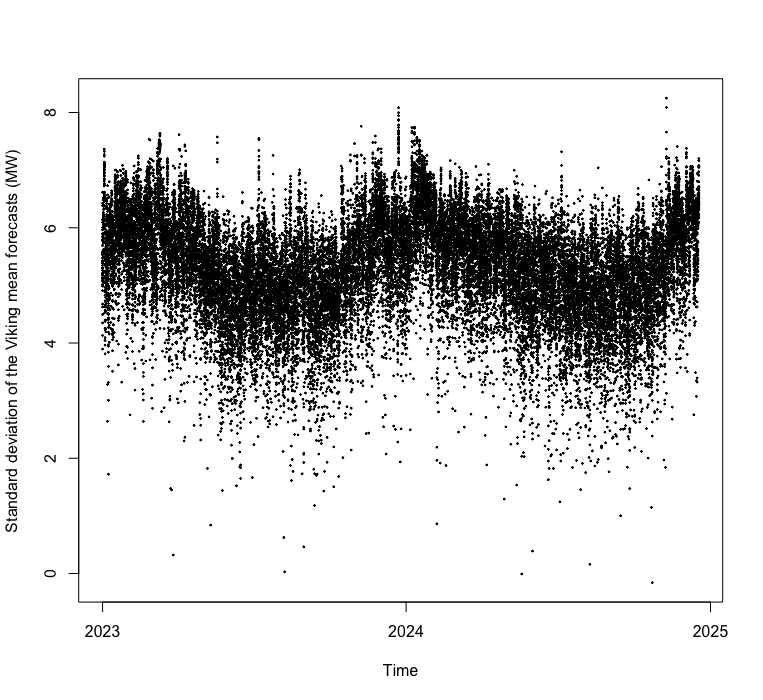}
\includegraphics[width=4.35cm]{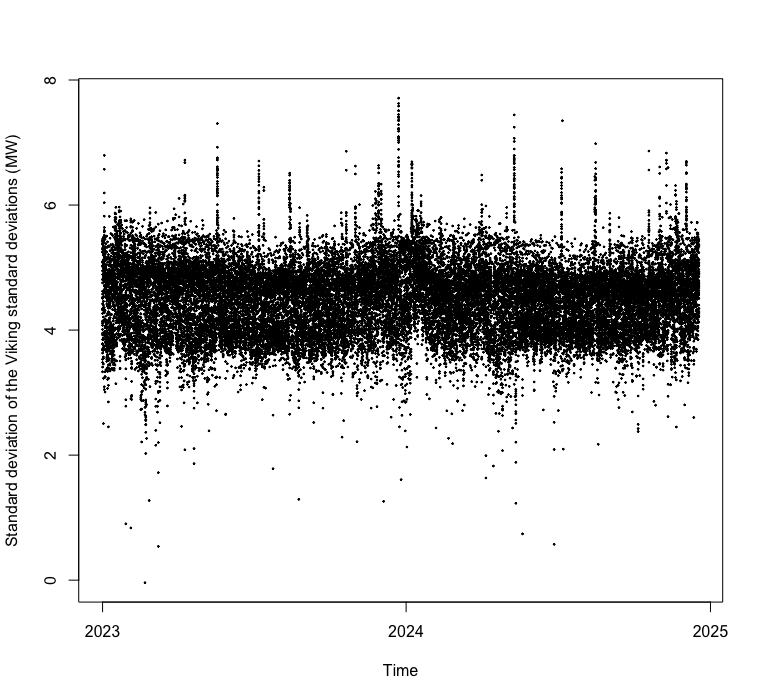}
\caption{Standard deviations of the Viking mean forecasts (on the left) and Viking standard deviations (on the right), used as dispersion variables in Gaulss Agg 2 and Gaulss Agg 3 respectively.}
\label{fig:sd}
\end{figure}

%
The other variants that utilize in two ways the dispersion of the standard deviations of the Viking algorithms do not capture the tails more precisely. We believe that the dispersion of the mean forecasts effectively captures divergence in meteorological forecasts and evolves much more over time. In contrast, the standard deviations of the Viking algorithms tend to be more stationary, as illustrated in Figure~\ref{fig:sd}.

\begin{figure}[t]
\centering
\includegraphics[width=4.35cm]{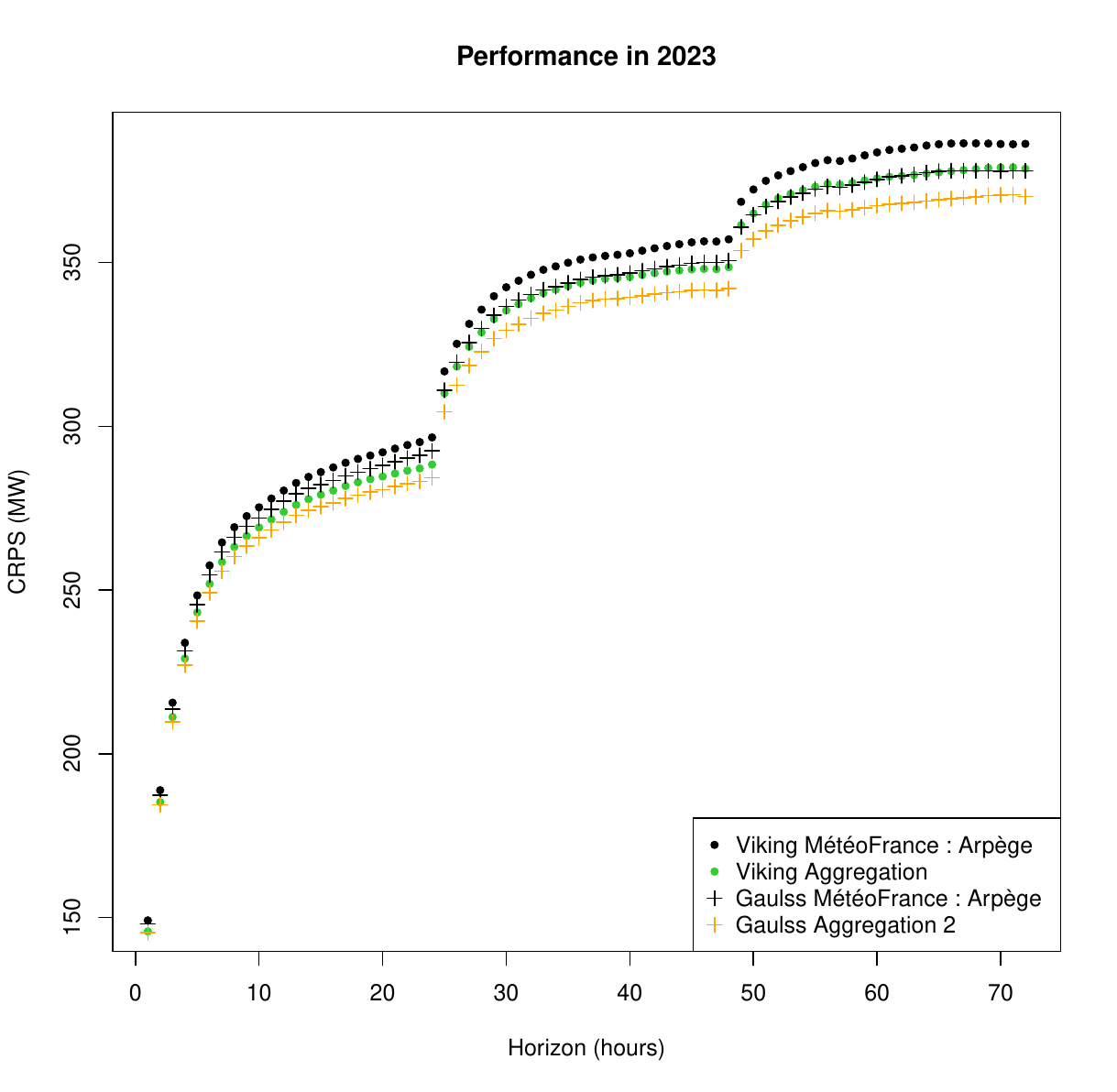}
\includegraphics[width=4.35cm]{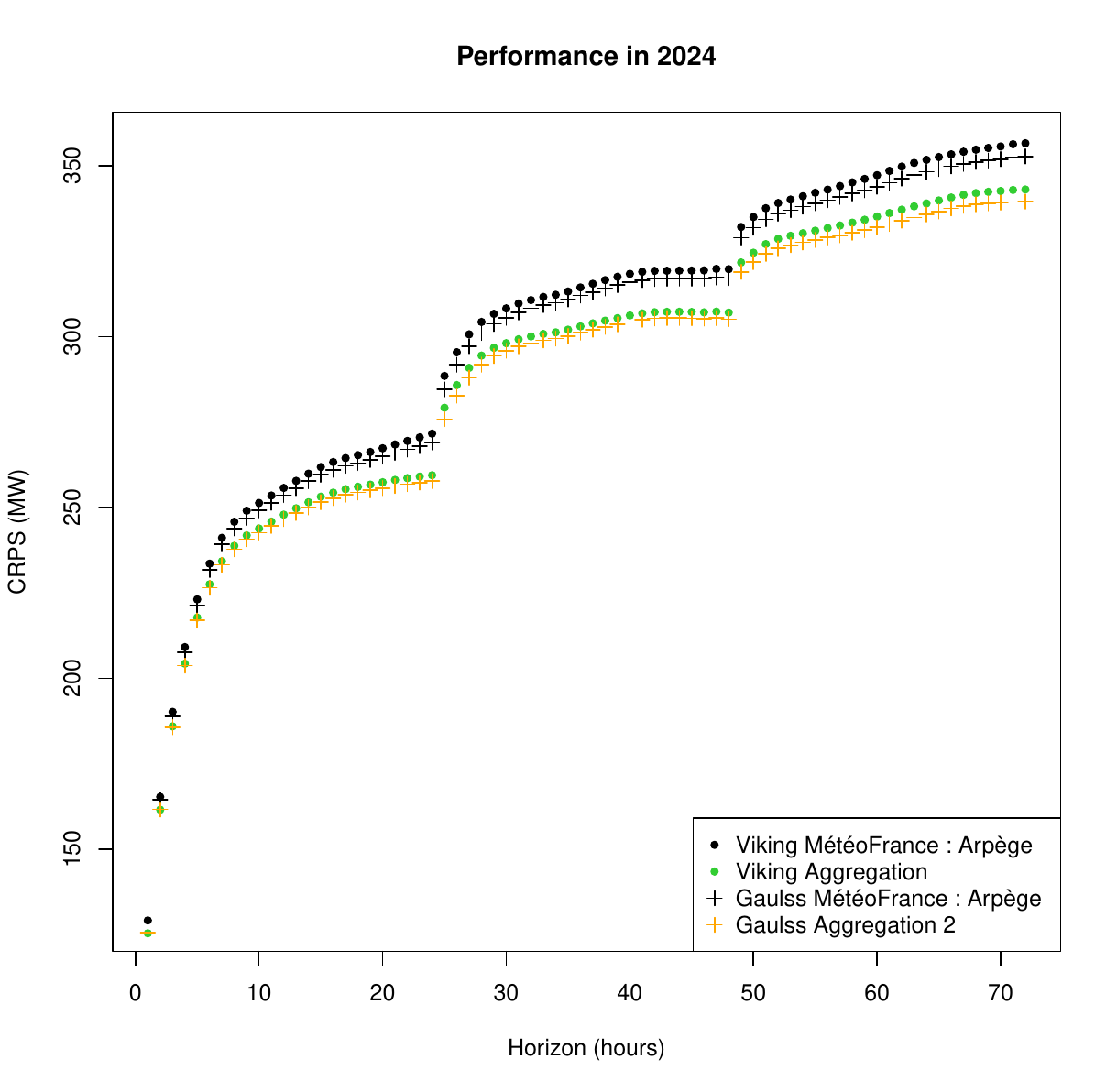}
\caption{CRPS of the aggregated probabilistic forecasts.}
\label{fig:crps_horizon}
\end{figure}

Finally, in Figure~\ref{fig:crps_horizon} the CRPS is displayed as a function of the forecast horizon. The short-term correction improves greatly the performances for very short-term horizons. As described in Section \ref{sec:evaluation}, the CRPS attributes low weights to the extreme quantiles, and the differences in CRPS between the considered probabilistic forecasts are low. Nonetheless, we observe an improvement due to the aggregation of meteorological forecast providers.

\section{Conclusion}
In this article, we developed a method to integrate multiple meteorological forecasts to enhance electricity load forecasting. Our approach involves applying an adaptive point forecasting method to each weather forecast provider, then aggregating the resulting point forecasts. Next, we calibrate a probabilistic model using the residuals from this improved point forecast, incorporating information about the initial forecasts’ variability. Finally, we implement a short-term correction.

An evaluation of the national electrical load in France found that aggregating forecasts from multiple meteorological providers yields better results than relying solely on the best individual provider. This conclusion holds for both point and probabilistic forecasting, with a greater improvement observed in the latter. The variability across multiple meteorological forecasts proves particularly beneficial for predicting the distribution of future targets, rather than just their average. 

A direction for future work is to make the method even more adaptive. In our current approach, the Gaulss coefficients and the short-term correction are fixed. We could learn the Gaulss coefficients adaptively, for example, using a gradient algorithm based on the log-likelihood. Additionally, the coefficient for the short-term correction might be adapted to respond more effectively to significant changes in the target. Another possibility is to adapt the quantile levels using adaptive conformal prediction (Zaffran et al., 2022).




\bibliographystyle{IEEEtran}
\bibliography{IEEEabrv,mybib.bib}

\end{document}